\documentclass[a4paper,9pt]{extarticle}
\usepackage{hyperref}
\usepackage{fullpage}
\usepackage[utf8x]{inputenc}
\usepackage{amssymb,amsmath}
\usepackage{floatrow}
\usepackage{graphicx}
\usepackage{wrapfig}
\usepackage{caption}
\usepackage{subcaption}
\usepackage{color}
\usepackage{multicol}
\usepackage{sidecap}
\usepackage{url}
\usepackage{color}

\newcommand\numberthis{\addtocounter{equation}{1}\tag{\theequation}}

\newenvironment{Figure}
  {\par\medskip\noindent\minipage{\linewidth}}
  {\endminipage\par\medskip}

\begin{document}

%\title{\textbf{Longitudinal coupling: A scalable scheme for superconducting qubits}}
\title{\textbf{Circuit design implementing longitudinal coupling: a scalable scheme for superconducting qubits}}
\author{Susanne Richer and David DiVincenzo\\
\textit{\small{JARA Institute for Quantum Information, RWTH Aachen University, D-52056 Aachen, Germany}}}
\maketitle

\renewcommand{\abstractname}{}
\begin{abstract}
We present a circuit construction for a fixed-frequency superconducting qubit and show how it can be scaled up to a grid with strictly local interactions. The circuit QED realization we propose implements $\sigma_z$-type coupling between a superconducting qubit and any number of $LC$ resonators. The resulting \textit{longitudinal coupling} is inherently different from the usual $\sigma_x$-type \textit{transverse coupling}, which is the one that has been most commonly used for superconducting qubits. In a grid of fixed-frequency qubits and resonators with a particular pattern of always-on interactions, coupling is strictly confined to nearest and next-nearest neighbor resonators; there is never any direct qubit-qubit coupling. We note that just a single unique qubit frequency suffices for the scalability of this scheme.  The same is true for the resonators, if the resonator-resonator coupling constants are varied instead. A controlled phase gate between two neighboring qubits can be realized with microwave drives on the qubits, without affecting the other qubits. This fact is a significant advantage for the scalability of this scheme.
\end{abstract}

\vspace{0.5cm}

\begin{multicols}{2}

\section{Introduction}
One of the most promising fields in quantum information is the implementation of circuit QED-based architectures using superconducting qubits. While qubit coherence times and coupling schemes have improved immensely in recent years \cite{chang}, \cite{barends}, \cite{yan}, the \textit{scalability} of these systems is still a huge problem.
We would like to implement a two-dimensional (2D) grid of qubits, where single-qubit gates and two-qubit gates between nearest-neighbor qubits can be done \textit{locally}, that is, without affecting the other qubits in the grid.
Most architectures either depend on tuning the qubits in and out of resonance to make gates \cite{barends} or on fixed-frequency qubits, where gates are made via microwave radiation \cite{blais}. While fixed-frequency qubits usually have larger coherence times than tunable qubits, scalability is constrained by the \textit{always-on interaction}. \\
However, Billangeon et al. present a scheme in \cite{bill} using a grid of fixed-frequency qubits and resonators with a particular pattern of always-on interactions that have strictly bounded range in the grid. When viewed in an appropriate frame, there is no coupling between any of the elements, unless we drive a certain two-qubit gate between nearest-neighbor qubits. As we will see, this architecture, which is based on what they refer to as \textit{longitudinal coupling}, has a high potential for scalability. 
Another useful application of longitudinal coupling is fast quantum non-demolition readout of a qubit via a resonator, as shown in \cite{didier}.\\ 
In this paper, we will start by giving a short summary of Billangeon's scheme in Sec. \ref{sec:trans_long}. We will then proceed by explaining what kind of Lagrangian we need for longitudinal coupling (Sec. \ref{sec:long}) and present a proposal for a circuit QED realization of a qubit-resonator unit with longitudinal coupling (Sec. \ref{sec:qubit}). 
Finally, in Secs. \ref{sec:n_resonators}-\ref{sec:grid}, we will explain how this idea can be scaled up to a grid with strictly local interactions. In the Appendices, we will analyze the effect of  unsymmetrical parameters due to imprecise fabrication, show how to do the transition from the Lagrangian to the Hamiltonian, and explain how the \textit{Cholesky transformation} can be used for variable elimination.

\section{Transverse and longitudinal coupling}
\label{sec:trans_long}

Let us first consider a system where a qubit is coupled to a resonator via its $\sigma_x$-degree of freedom. Most circuit QED structures result in this kind of coupling, which we will denominate \textit{transverse coupling}, in contrast to the $\sigma_z$-type \textit{longitudinal coupling} we will study below. The corresponding Hamiltonian is the \textit{Rabi Hamiltonian}

\begin{align}
\mathcal{H} = \omega_r a^\dagger a + \frac{\Delta}{2} \sigma_z + g \,  \sigma_x (a^\dagger + a),
\label{eq:Rabi}
\end{align}

where $\omega_r$ is the frequency of the resonator and $\Delta$ is the gap of the qubit, which is taken to be a two-level system. \\
If the coupling $g$ between the qubit and the resonator is small compared to their detuning $\Delta - \omega_r$ (\textit{dispersive regime}),  we can approximately diagonalize the Hamiltonian using the \textit{Schrieffer-Wolff} unitary transformation 

\begin{align}
\mathcal{U} = \exp{\left(\gamma (a^\dagger \sigma_- - a \sigma_+) - \bar{\gamma} (a^\dagger \sigma_+ - a \sigma_-)\right)}
\label{eq:SW}
\end{align}

with $\gamma = g/(\Delta - \omega_r)$, and $\bar\gamma = g/(\Delta + \omega_r)$ (see \cite{winkler}, \cite{zueco} and \cite{ich}). Note that this is a perturbative treatment, where to second order in $g$ we find 

\begin{align}
\mathcal{H}' = \omega_r a^\dagger a + \frac{\Delta}{2} \sigma_z + \chi \, \sigma_z (a^\dagger + a)^2 
\label{eq:dispersive}
\end{align}

with $\chi = g\,(\gamma + \bar\gamma)/2$. The last term in Eq. \ref{eq:dispersive} is the so-called \textit{dispersive shift} that makes the transition frequency of the resonator dependent on the qubit's state and vice versa. While this shift is useful for read-out, it also means that the always-on interaction between the resonator and the qubit will inevitably entangle them. Unlike for the longitudinal coupling (see below), there is no local frame in which the coupling is turned off.
Another undesirable effect of the dispersive shift is that it makes the qubit relaxation rate dependent on the photon lifetime (the so-called \textit{Purcell effect} \cite{purcell}).\\
If we couple two qubits to the same resonator and use it as a quantum bus, a perturbative transformation similar to Eq. \ref{eq:SW} will indicate direct always-on coupling between the qubits.
This always-on coupling can be problematic when we want to address qubits separately in a larger grid.
\\
\\
Mindful of these problems, Billangeon et al. proposed a scheme in \cite{bill} using \textit{longitudinal coupling}, that is, a system with the Hamiltonian

\begin{align}
\mathcal{H} = \omega_r \, a^\dagger a + \frac{\Delta}{2} \,\sigma_z + g \, \sigma_z (a^\dagger + a),
\label{eq:long}
\end{align}

where the qubit couples to the resonator via a $\sigma_z$-operator, instead of the usual (\textit{transverse}) $\sigma_x$-type coupling of the Rabi Hamiltonian (Eq. \ref{eq:Rabi}). Using the unitary transformation (\textit{Lang-Firsov transformation} \cite{firsov}) 

\begin{align}
U = e^{-\theta \sigma_z (a^\dagger - a)}
\label{eq:lang-firsov}
\end{align}

with $\theta= g/\omega_r$, the Hamiltonian can be exactly diagonalized, showing that there is no dispersive shift:

\begin{align}
\mathcal{H}' = \omega_r \, a^\dagger a + \frac{\Delta}{2} \sigma_z - \frac{g^2}{\omega_r} \mathbf{1}.
\label{eq:long_result}
\end{align}

Remarkably, the energies associated with the qubit and the resonator are unaffected by the longitudinal coupling.
Note that while the treatment in the transverse case (Eq. \ref{eq:SW}) is only perturbative, the Lang-Firsov transformation is \textit{exact}, with no restrictions on the coupling strength or detuning of the system. This has the advantage that even for large coupling strengths, the qubit relaxation rate is not degraded by a dependence on the finite photon lifetime. We have thus gone to a frame where the Hamiltonian is diagonal without any residual coupling.\\
A transverse drive on the qubit 

\begin{align}
\mathcal{H}_d (t) = \Omega \cos(\omega t + \phi) \sigma_x
\label{eq:drive}
\end{align}

has to be transformed to the same frame, again using Eq. \ref{eq:lang-firsov}. The interaction induced by this drive will be rigorously confined to a small neighborhood of the qubit being driven, without generating any direct qubit-qubit coupling (see below). As shown explicitly in \cite{bill}, a drive at the qubit's frequency $\omega = \Delta$, enables us to do single-qubit operations (within the rotating wave approximation (RWA)), while a drive at $\omega = |\Delta \pm k \, \omega_r|$ leads to sideband transitions of order $k$ between the qubit and the resonator that will later be used to implement a C-phase gate between neighboring qubits.\\
These sideband transitions are possible due to the absence of the dispersive shift in Eq. \ref{eq:long_result}, since a certain transition will stay resonant irrespective of the number of photons in the resonators. 

\begin{Figure}
  \begin{center}
    \includegraphics[width=.85\linewidth]{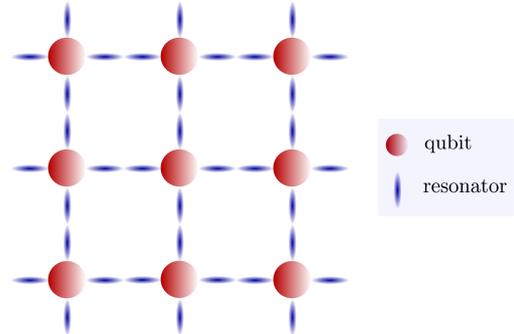}
    \captionof{figure}{In \cite{bill}, a 2D lattice of qubits is proposed in which each qubit couples to four resonators via its longitudinal degree of freedom and every resonator couples to a resonator of the next unit cell via an orthogonal degree of freedom. Coupling is strictly restricted to the nearest and next-nearest neighbor resonators of each qubit. In particular, there is never any direct qubit-qubit coupling. Therefore, all qubits could, in principle, have the same frequency. The same is true for the resonators, as shown in Appendix \ref{app:diagonal}.}  
    \label{grid.pdf}
  \end{center}
\end{Figure}

This idea can be extended to a grid (see Fig. \ref{grid.pdf}), where a unit cell consists of a qubit coupled longitudinally to four resonators and every resonator is coupled to a resonator of the neighboring unit cell. As shown in \cite{bill}, the associated Hamiltonian can be exactly diagonalized  if the two resonators are coupled through an orthogonal degree of freedom (that is, $i (a^\dagger - a)$ instead of $(a^\dagger + a)$).\\ 
The desired Hamiltonian for two qubits coupled via two resonators is

\begin{align*}
\mathcal{H} = \sum_{i=1}^2 & \omega_i\, a_i^\dagger a_i +  \frac{\Delta_i}{2} \, \sigma_{i}^z + g_i \, \sigma_i^z \, (a_i^\dagger + a_i)  \\
&- g_c \,(a_1^\dagger - a_1)(a_2^\dagger - a_2). \numberthis
\label{eq:long2}
\end{align*}

Given that this system is exactly diagonalizable, and there exists a frame without dispersive shifts or residual couplings, driving one qubit has no effect on the neighboring qubits, as the interaction is strictly confined to the nearest and next-nearest neighbor resonators. There is rigorously no qubit-qubit interaction. By choosing the frequency of the transverse drive on the qubit (Eq. \ref{eq:drive}), sideband transitions can be driven either between the qubit and a nearest neighbor resonator or between the qubit and both the nearest and next-nearest neighbor resonators (within the RWA).As coupling will be strictly restricted to the nearest and next-nearest neighbor resonators of each qubit, all qubits could, in principle, have the same frequency. As shown in Appendix \ref{app:diagonal}, the same is true for the resonators. While the qubit frequencies are unaffected by the diagonalization of $\mathcal{H}$, the resonator frequencies are shifted (see Eq. (B9) in \cite{bill} and Appendix \ref{app:diagonal}). By varying the coupling between the resonators, we can ensure that all eight resonators that surround a qubit have a different bare frequency, even if all eight original frequencies were equal. This allows us to choose unambiguously which sideband transition we want to drive (see Fig. \ref{grid.pdf}). \\
Read-out and a controlled-phase gate between two neighboring qubits are possible via a series of sideband transitions between either qubit and one or both resonators. There is never any direct qubit-qubit coupling needed. The fact that all other qubits are unaffected by these actions is a significant advantage concerning the scalability of this scheme. \\
We want to realize the idea presented in \cite{bill} using superconducting qubits. While there are two proposals for such a qubit-resonator system in \cite{bill}, numerical calculations are needed to determine the type of coupling. Our approach, in contrast, uses a very simple one-junction qubit, where the coupling can be characterized with purely analytic considerations. Whereas other implementations of longitudinal coupling in circuit QED (\cite{kerman}, \cite{braumueller}) rely on changing the qubit's resonance frequency, our approach is inherently different, as it exploits the parity of the interaction term while keeping the resonance frequency fixed.
In the circuit we propose each qubit couples longitudinally to four resonators via the phase degree of freedom and each resonator couples to the next one via the charge degree of freedom. This proposal is easily scalable to any number of resonators per qubit and any number of 1-qubit-n-resonator unit cells, as we will show in Secs. \ref{sec:n_resonators}-\ref{sec:grid}.

\section{What is longitudinal coupling?}
\label{sec:long}

How do we determine whether a coupling term in an electric circuit is longitudinal or transverse? While the $\sigma_x$-operator of a qubit couples the qubit basis states $\langle 0| \sigma_x |1\rangle \neq 0$, the $\sigma_z$-operator couples the states $|\pm\rangle = (|0\rangle \pm |1\rangle)/\sqrt{2}$, that is $\langle +| \sigma_z |-\rangle \neq 0$.\\
Consider a system of a qubit coupled to a resonator via a phase degree of freedom (see Sec. \ref{sec:qubit}). Suppose we can write a coupling term between qubit and resonator as a product function $f(\varphi_q,\varphi_r) = f_q(\varphi_q) f_r(\varphi_r)$ of the phase variables $\varphi_q$ for the qubit and $\varphi_r$ for the resonator. In this picture, the qubit states correspond to wavefunctions and $\sigma_z$-type coupling is given by the matrix element

\begin{align}
\langle+|g \,& \sigma_z|-\rangle 
= \int_{-\infty}^\infty \psi_{+}(\varphi_q) f_q(\varphi_q)  \psi_{-}(\varphi_q) \, d\varphi_q.
\end{align}

For this to be nonzero, we need $f_q \, \psi_+ \psi_-$ to be an even function of $\varphi_q$ (as the integral of an odd function over a symmetric interval is always zero).\\ 
The appropriate wavefunctions for approximately harmonic potentials are superpositions of harmonic oscillator wavefunctions, where $\psi_0$ is an even function in $\varphi_q$ and $\psi_1$ is odd. Knowing that $\psi_+ \psi_- = (\psi_0^2 - \psi_1^2)/2$ is an even function (see Fig. \ref{wavefunctions.pdf}), $f_q$ must be also for nonzero longitudinal coupling. On the other hand, a coupling term that is an odd function in $\varphi_q$ will give transverse coupling, as $\psi_0 \psi_1$ is odd.

\begin{Figure}
  \begin{center}
    \includegraphics[width=.85\linewidth]{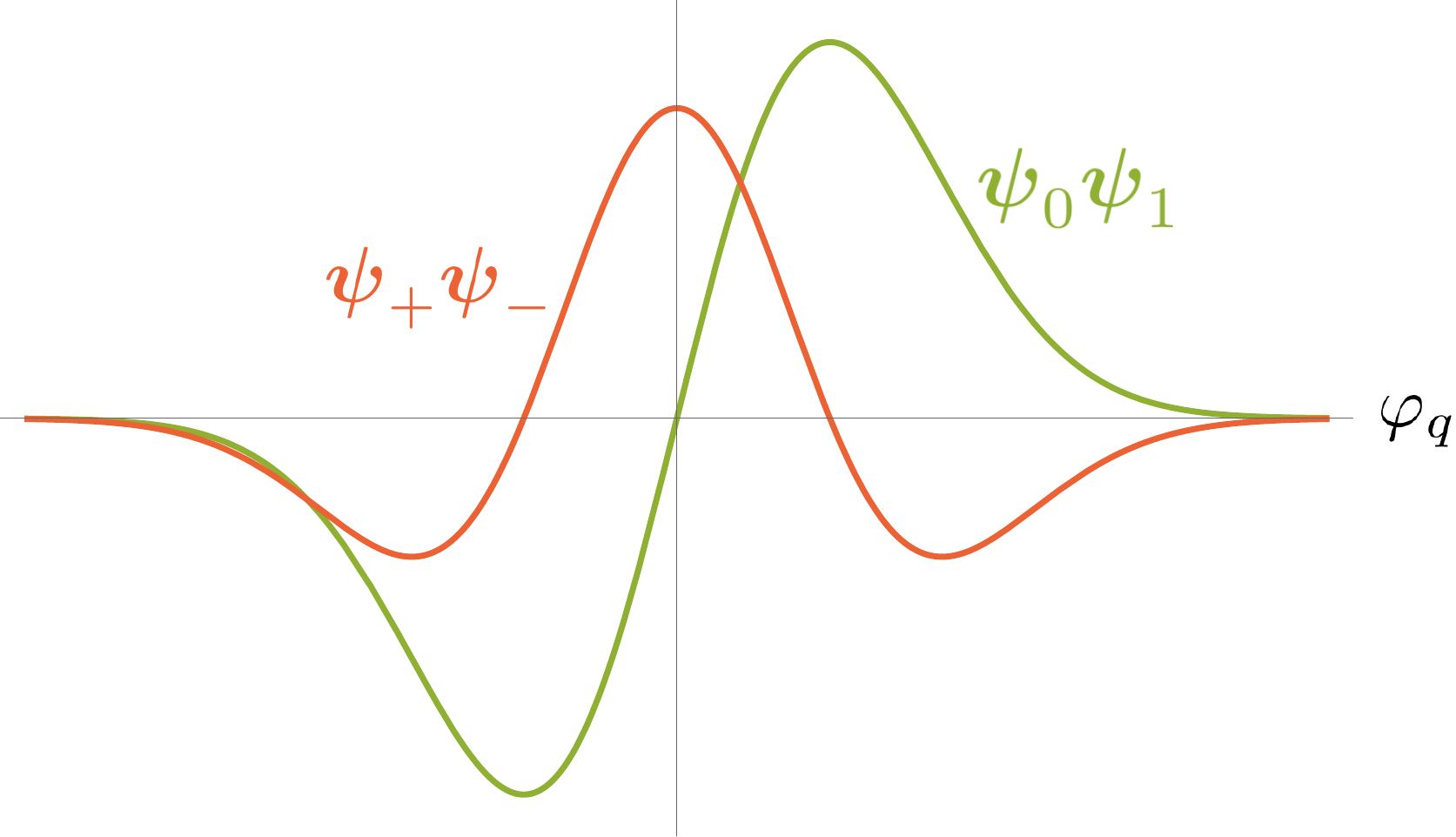}
    \captionof{figure}{While the product of the wavefunctions $\psi_+$ and $\psi_-$ (orange curve) is an even function in $\varphi_q$, $\psi_0\psi_1$ (green curve) is an odd function.}
    \label{wavefunctions.pdf}
  \end{center}
\end{Figure}

Similar reasoning for the resonator variable leads to the conclusion that the coupling term $g \, \sigma_z (a^\dagger + a)$ corresponds to a function that is \textit{even in} $\varphi_q$ and \textit{odd in} $\varphi_r$. Our objective is to create electric circuits with this form of coupling.

\section{Qubit coupled to resonator}
\label{sec:qubit}
Throughout this paper, we will describe superconducting circuits in terms of nodes and branches, where every circuit element is a branch that connects two nodes. We will take the superconducting phases at the nodes as the variables (following \cite{devoret}), where the phase at a node is the rescaled flux or, equivalently, the time integral over the node voltage:

\begin{align}
\varphi_i = \left(\frac{2\pi}{\Phi_0}\right) \Phi_i = \left(\frac{2\pi}{\Phi_0}\right) \int_{-\infty}^t V_i(t') \, dt'.
\end{align}

\begin{wrapfigure}{r}{0.58\linewidth}
\vspace{-20pt}
  \begin{center}
    \includegraphics[width=\linewidth]{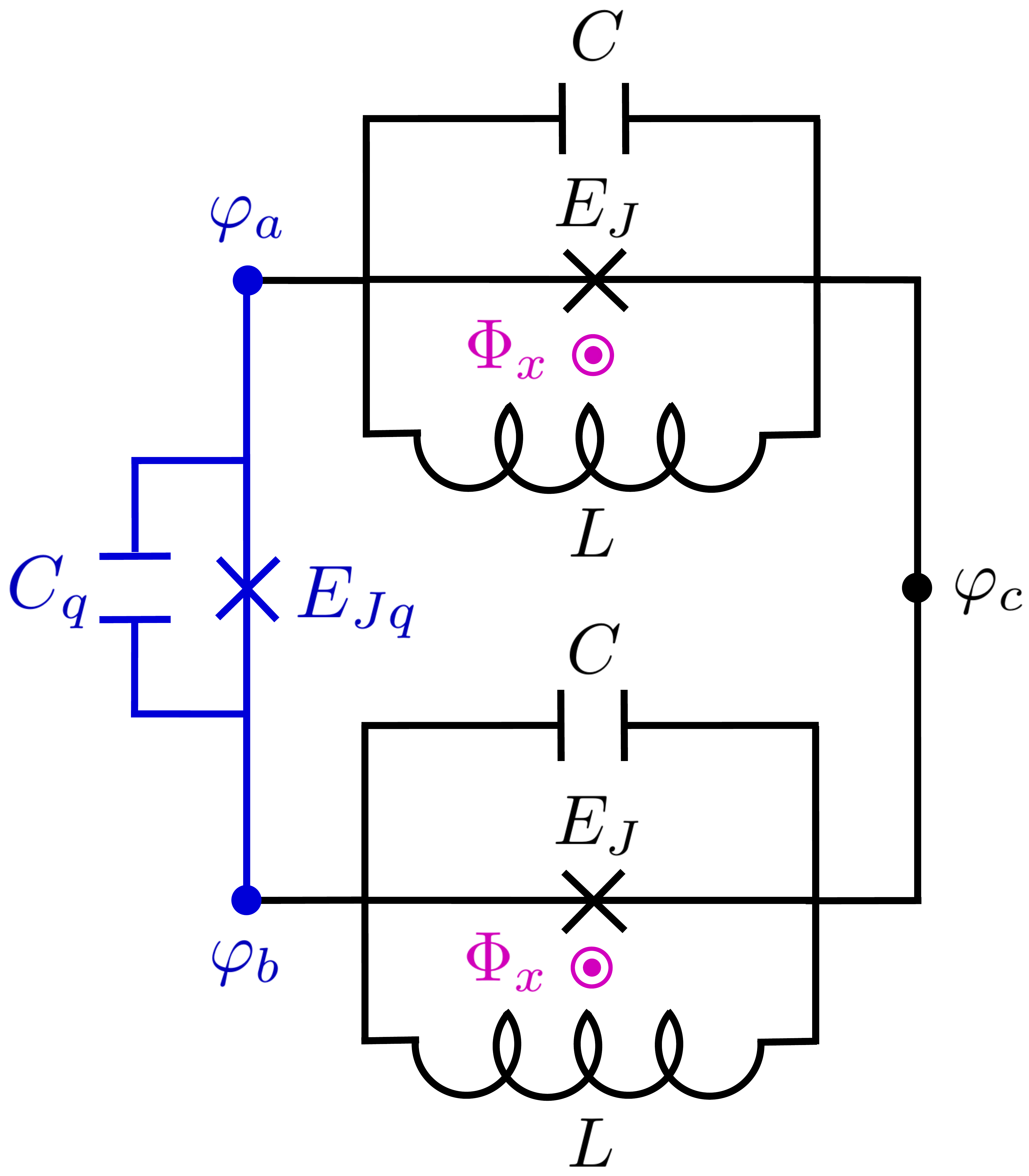}
  \end{center}
  \vspace{-10pt}
   \caption{Qubit (blue junction) coupled to a resonator via its longitudinal degree of freedom.}
    \label{qubit_resonator.pdf}
\end{wrapfigure}

Note that, as the overall phase is undefined, the real variables will be the phase differences between the nodes, which reduces the number of independent variables by 1. \\
Fig. \ref{qubit_resonator.pdf} depicts our central concept for a lumped element system of a qubit coupled to a resonator via its longitudinal degree of freedom. The qubit is the simplest possible design: a single junction with a bare capacitance (depicted in blue).
We assume that the upper and lower branches in Fig. \ref{qubit_resonator.pdf} are identical (for the unsymmetrical case see Appendix \ref{app:unequal}) and that the loops created by the inductances and the Josephson junctions are each threaded by a flux $\Phi_x=\Phi_0/4$. 
Writing the Lagrangian in terms of the node phases yields

\begin{align*}
\mathcal L &=   \left(\frac{\Phi_0}{2\pi}\right)^2 \left(\frac{C}{2} (\dot\varphi_a-\dot\varphi_c)^2 + \frac{C}{2}(\dot\varphi_b-\dot\varphi_c)^2 \right. \\ 
& \left. + \frac{C_q}{2} (\dot\varphi_a-\dot\varphi_b)^2 - \frac{1}{2L} (\varphi_a - \varphi_c)^2 - \frac{1}{2L} (\varphi_b - \varphi_c)^2 \right) \\
&+ E_{Jq} \cos(\varphi_a - \varphi_b)+E_J \sin(\varphi_a - \varphi_c) \\
&+E_J \sin(\varphi_b - \varphi_c), \numberthis
\end{align*}

where the threading fluxes turn the last two terms from cosine to sine. We introduce 

\begin{align}
\varphi_q = \varphi_a - \varphi_b, \qquad \varphi_p = \varphi_a + \varphi_b
\end{align}

and rewrite the Lagrangian in these new variables

\begin{align*}
\mathcal L &= \left(\frac{\Phi_0}{2\pi}\right)^2 \bigg(\frac{2\,C_q+C}{4} \, \dot\varphi_q^2 + \frac{C}{4} \, (\dot\varphi_p - 2\,\dot \varphi_c)^2 \\ 
&  - \frac{1}{4L} \,\left(\varphi_q^2 + (\varphi_p - 2\,\varphi_c)^2\right) \bigg) + E_{Jq} \cos(\varphi_q) \\
&+ 2 E_J \cos\left(\frac{\varphi_q}2\right) \sin\left(\frac{\varphi_p - 2\,\varphi_c}2\right). \numberthis
\label{eq:lagr}
\end{align*}

Obviously, there are only two independent variables in Eq. \ref{eq:lagr}, namely $\varphi_q$ and

\begin{align}
 \varphi_r = \varphi_p - 2\,\varphi_c.
\end{align}

We call $\varphi_q$ the qubit variable as its potential consists of an anharmonic Josephson term and an additional harmonic term. The second variable $\varphi_r = \varphi_a + \varphi_b - 2\,\varphi_c$ has a \textit{purely} harmonic potential (apart from the coupling term) and will therefore serve as the resonator. \\ 
With this choice of variables, all coupling terms between the qubit and the resonator via the capacitances and inductances cancel out completely and the only coupling term that is left is the desired longitudinal one

\begin{align*}
\mathcal L &= \left(\frac{\Phi_0}{2\pi}\right)^2 \left(\frac{2\,C_q+C}{4} \, \dot\varphi_q^2 + \frac{C}{4} \, \dot\varphi_r^2 - \frac{1}{4L} \, (\varphi_q^2 + \varphi_r^2) \right) \\
&+ E_{Jq} \cos(\varphi_q) + 2E_J \cos\left(\frac{\varphi_q}2\right) \sin\left(\frac{\varphi_r}2\right). \numberthis
\label{eq:lagr_final}
\end{align*}

\begin{wrapfigure}{r}{0.65\linewidth}
\vspace{-30pt}
  \begin{center}
    \includegraphics[width=.9\linewidth]{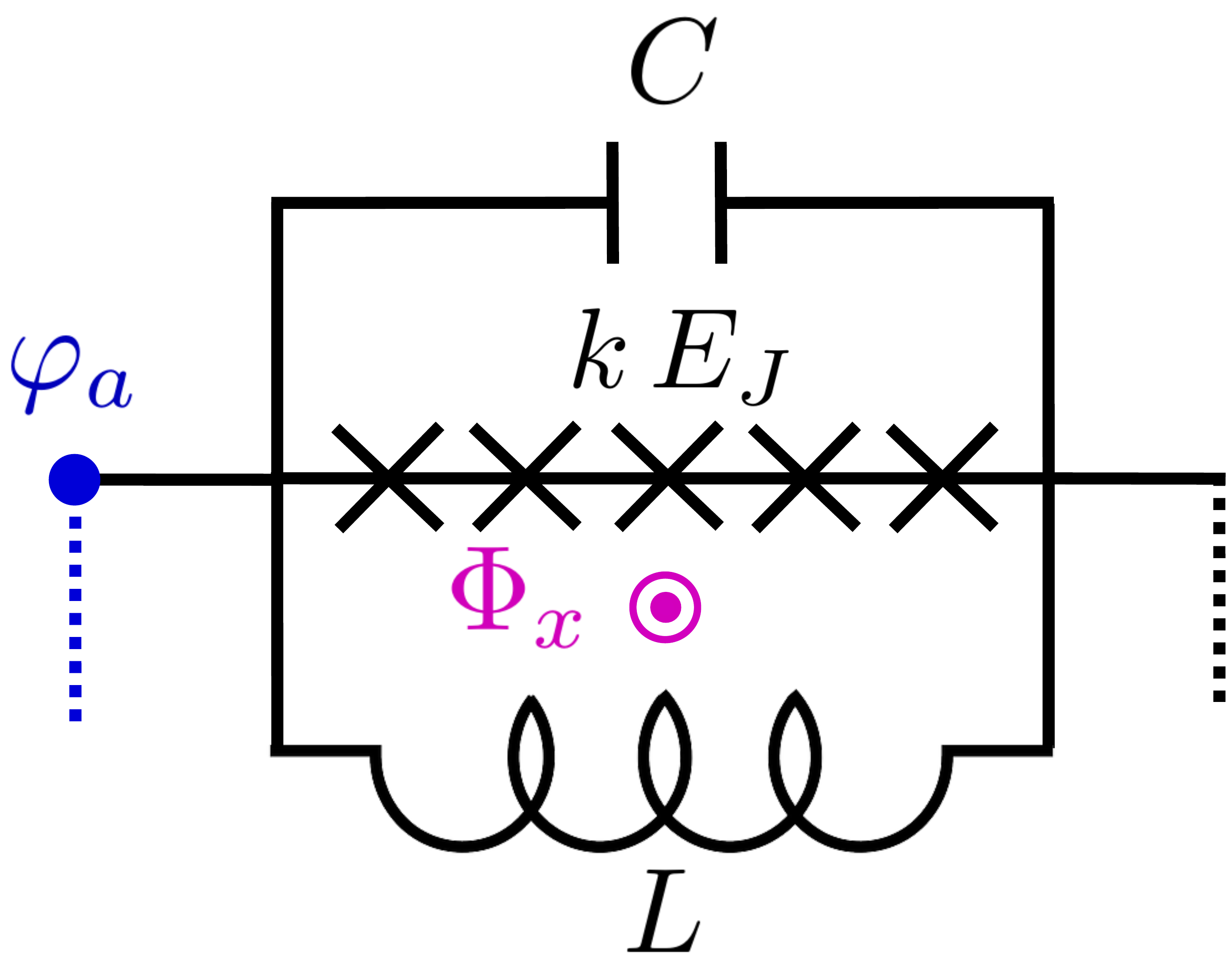}
  \end{center}
  \vspace{-10pt}
   \caption{Multiple coupling junctions. The flux has to be adjusted to $\Phi_x = k \, \Phi_0/4$.}
    \label{many_junctions.pdf}
\end{wrapfigure}

The final term of $\cal L$ is a coupling term that is even in the qubit variable and odd in the resonator variable; as discussed in Sec. \ref{sec:long}, this gives the coupling the longitudinal form.
Observe that the qubit-resonator system is completely uncoupled for $E_J = 0$. \\
Now assuming the qubit's potential is anharmonic enough to justify a two-level approximation, we find the coupling term to be

\begin{align*}
2E_J &\cos\left(\frac{\varphi_q}2\right) \sin\left(\frac{\varphi_r}2\right) \\ &\to g \, \sigma_z \left((a^\dagger + a) + \mathcal{O}((a^\dagger + a)^3) \right) \numberthis
\end{align*}
(compare Sec. \ref{sec:long}). If we want to get rid of the higher order resonator terms, we could substitute the coupling junctions by a row of $k$ equal junctions and adjust the flux to $\Phi_x = k \, \Phi_0/4$ as depicted in Fig. \ref{many_junctions.pdf}. Such Josephson arrays are oftentimes used in fluxonium qubits (see \cite{fluxonium}). As described in detail in \cite{viola}, the collective modes generated by the geometric capacitances of these junctions can be ignored as they are of much higher energy. Therefore, the use of a Josephson array leads in a simple way to a linearized coupling term

\begin{align*}
2 E_J \cos(\varphi_q/(2k)) \sin(\varphi_r/(2k)) & \approx E_J \, \varphi_q^2 \, \varphi_r \\ & \to g \, \sigma_z (a^\dag + a),\numberthis
\label{eq:series}
\end{align*}

which corresponds exactly to the longitudinal coupling term in Eq. \ref{eq:long}.
This series expansion is the only approximation ever made in the derivation of this Lagrangian.

Note that we do not need to consider charge offsets here, as all charge islands are short-circuited by the inductors (see Fig. \ref{qubit_resonator.pdf} and \cite{fluxonium}). Charge fluctuations are a big problem for superconducting qubits, such as the Cooper pair box, which can be reduced by choosing certain parameter ranges (see \cite{koch} on transmon qubits). A circuit such as ours that is inherently free of charge-offsets circumvents any such restrictions on the parameter ranges.

\section{Extension to n resonators}
\label{sec:n_resonators}

The idea presented in Sec. \ref{sec:qubit} is easily extendable to a block of one qubit coupled separately to any number of resonators as depicted in Fig. \ref{qubit_many_resonators.pdf}. 
Fortunately, adding another resonator arm to the system does not have any effect on the first resonator and the only coupling terms are the ones between the qubit and each resonator

\begin{align*}
\mathcal L &= \sum_{j} \left(\frac{\Phi_0}{2\pi}\right)^2 \left(\frac{2\,C_q+C_j}{4} \, \dot\varphi_q^2 + \frac{C_j}{4} \, \dot\varphi_{r,j}^2 \right. \\& \left.
- \frac{1}{4L_j} \, (\varphi_q^2 + \varphi_{r,j}^2) \right) + E_{Jq} \cos(\varphi_q) \\
&+ 2E_{J,j} \cos\left(\frac{\varphi_q}2\right) \sin\left(\frac{\varphi_{r,j}}2\right), \numberthis
\end{align*}

where again

\begin{align}
\varphi_q = \varphi_a - \varphi_b \quad \text{and} \quad \varphi_{r,j} = \varphi_a + \varphi_b - 2 \,\varphi_{c,j}.
\end{align}

While we need the upper and lower branch of each resonator to be equal (see Sec. \ref{sec:qubit} and Appendix \ref{app:unequal}), the individual resonator arms are completely independent of each other and can have different capacitances, inductances and Josephson junctions (that is, different resonator frequencies and coupling constants).

\begin{Figure}
  \begin{center}
    \includegraphics[width=.8\linewidth]{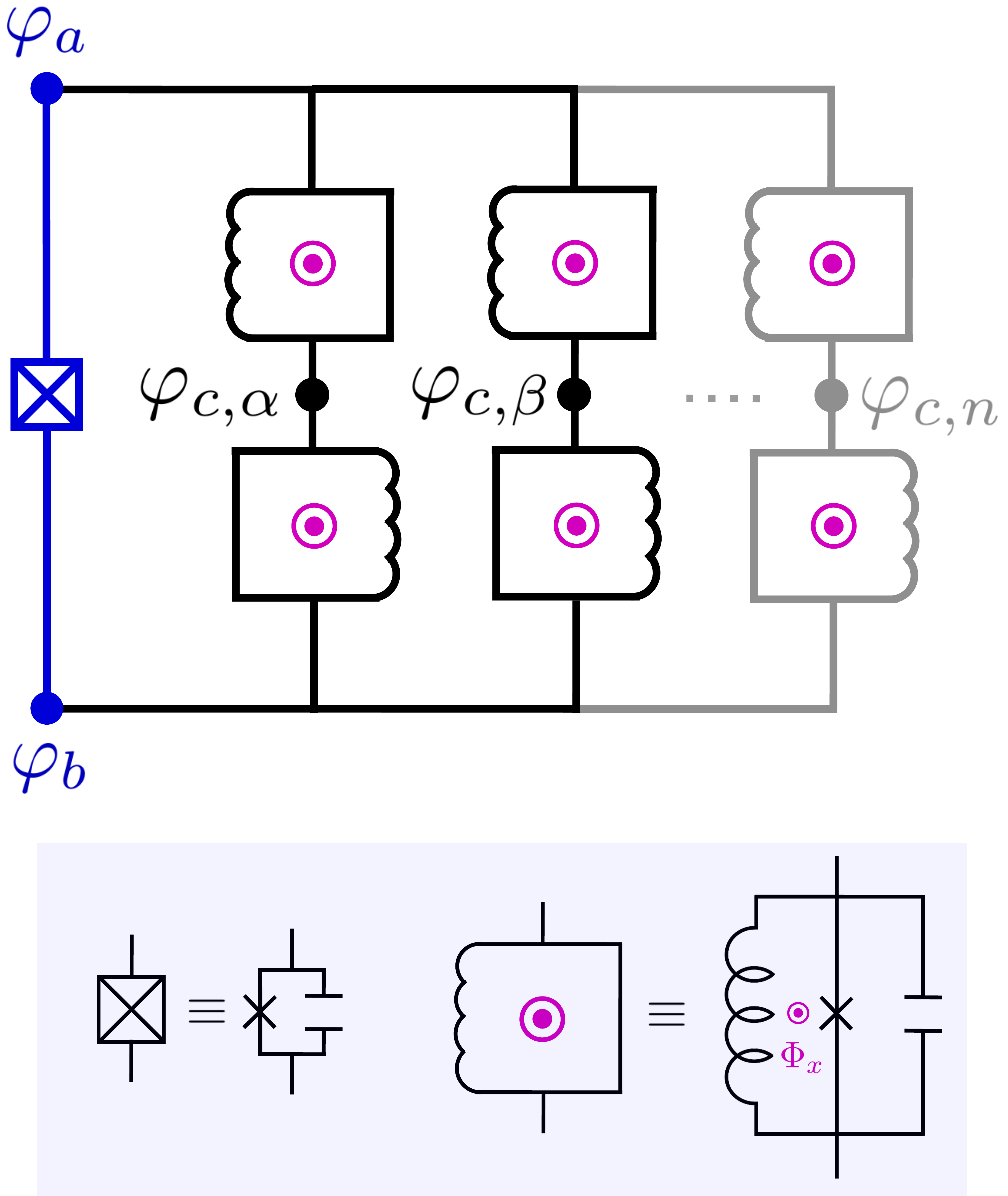}
    \captionof{figure}{Qubit coupled independently to $n$ resonators via its longitudinal degree of freedom. We do not show compensating fluxes here; note that in the larger grid implementation (Fig. \ref{ring.pdf}) all larger loops in the circuit will always contain whole numbers of flux quanta, so no compensating fluxes are needed there.} 
    \label{qubit_many_resonators.pdf}
  \end{center}
\end{Figure}

As each resonator arm adds another harmonic term to the qubit's potential, it will be important to set $E_{Jq} $ to be sufficiently large compared with $(\Phi_0/(2\pi))^2 \sum_j 1/(4L_j)$ to maintain the qubit's anharmonicity (see appendix \ref{app:hamiltonian}).

\section{Two coupled blocks}
\label{sec:two_blocks}

Now we want to couple one of these blocks to the next one via the charge degree of freedom of the resonator in order to implement Eq. \ref{eq:long2}. Fig. \ref{two_blocks.pdf} shows the simplest way we have found to obtain the desired coupling between two blocks.

\begin{Figure}
  \begin{center}
    \includegraphics[width=\linewidth]{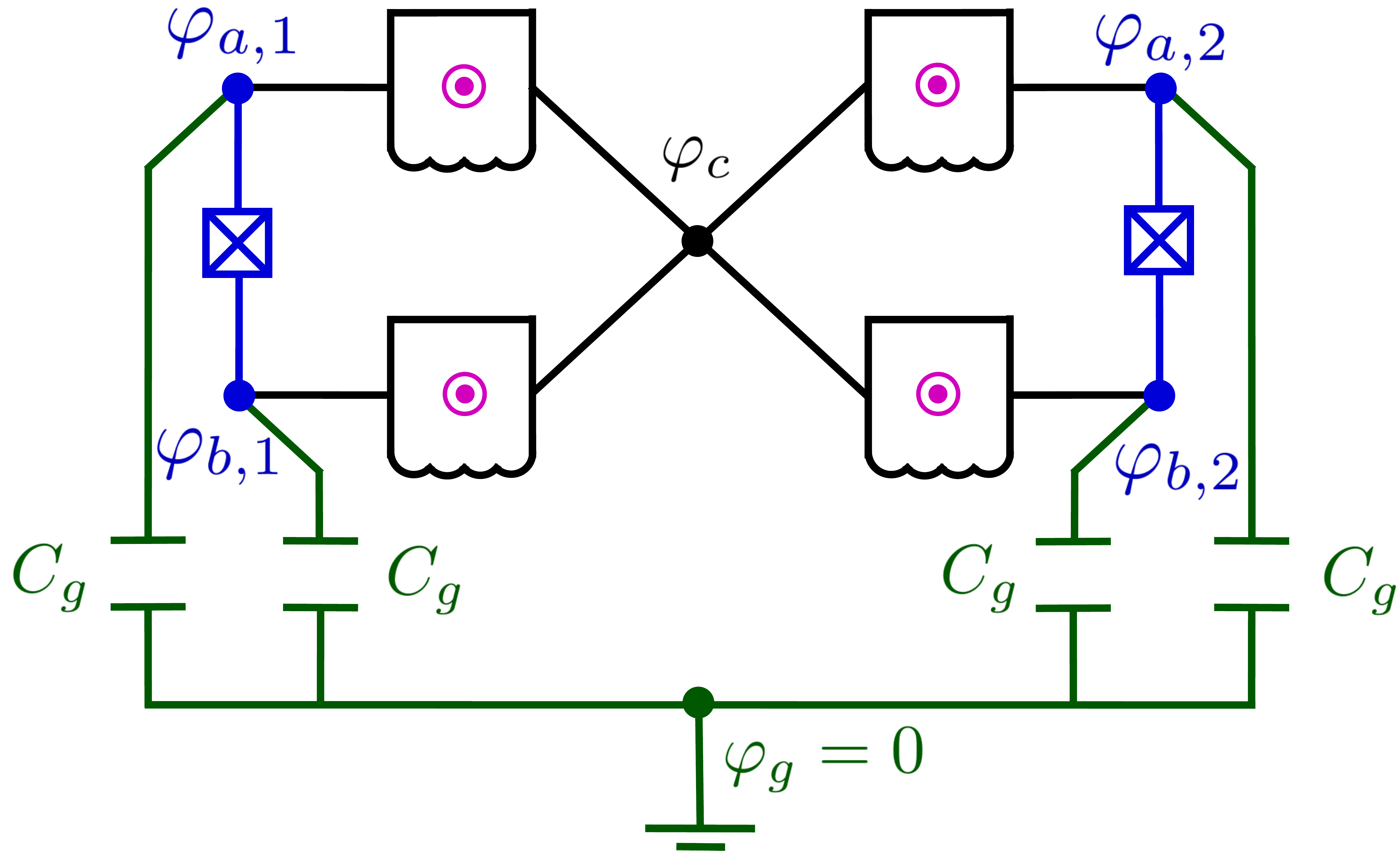}
    \captionof{figure}{This circuit realizes two qubits and two resonators coupled according to the scheme of \cite{bill}.}
    \label{two_blocks.pdf}
  \end{center}
\end{Figure}

We connect two neighboring resonator branches by tying their $\varphi_{c}$ nodes together and connect all qubit nodes to a common ground node $\varphi_g$ via capacitances $C_g$. Note that the two blocks are uncoupled for $C_g = 0$. As the node phases are only defined up to an overall phase, we have the freedom to set $\varphi_g = 0$ (\textit{ground node} \cite{devoret}) without loss of generality. However, we will keep $\varphi_g$ here in order to show that possible charge offsets between ground and the rest of the circuit can not influence our system.
The kinetic energy of this coupled system is

\begin{align*}
\mathcal T &=  \sum_{i=1}^2 \left(\frac{\Phi_0}{2\pi}\right)^2 \biggl(\frac{C_{q,i}}{2} (\dot\varphi_{a,i} - \dot\varphi_{b,i})^2 + \frac{C_i}{2} (\dot\varphi_{a,i} - \dot\varphi_{c})^2 \\
&+ \frac{C_i}{2} (\dot\varphi_{b,i} - \dot\varphi_{c})^2 +\frac{C_g}{2} \left( (\dot\varphi_{a,i} - \dot\varphi_g)^2 + (\dot\varphi_{b,i} - \dot\varphi_g)^2 \right) \biggr) \\
&= \sum_{i=1}^2  \left(\frac{\Phi_0}{2\pi}\right)^2 \biggl(\frac{C_{q,i}}{2} \dot\varphi_{q,i}^2 + \frac{C_i + C_g}4 \, (\dot\varphi_{q,i}^2 + \dot\varphi_{r,i}^2)   \\
&+ 2\,C_g \, (\dot\varphi_{c} - \dot\varphi_g)^2 + C_g \, (\dot\varphi_{c} - \dot\varphi_g) \, (\dot\varphi_{r,1} + \dot\varphi_{r,2}) \biggr), \numberthis
\label{eq:T_two_blocks}
\end{align*}

again with

\begin{align*}
\varphi_{q,i} = \varphi_{a,i} - \varphi_{b,i}, \qquad 
\varphi_{r,i} = \varphi_{a,i} + \varphi_{b,i} - 2 \,\varphi_{c}. \numberthis
\label{eq:variables}
\end{align*}

We can see that, apart from the two qubit and resonator variables, a fifth variable $\varphi_{c} - \varphi_g$ appears that mediates the coupling between the resonator variables $\varphi_{r,1}$ and $\varphi_{r,2}$. Introducing

\begin{align}
\varphi_* = \varphi_c - \varphi_g + \frac{\varphi_{r,1} + \varphi_{r,2}}{4}
\label{eq:eliminate}
\end{align}

leads to

\begin{align*}
\mathcal T = \sum_{i=1}^2 & \left(\frac{\Phi_0}{2\pi}\right)^2 \biggl(\frac{2\, C_{q,i} + C_i + C_g}4 \, \dot\varphi_{q,i}^2 + \frac{C_i}4 \, \dot\varphi_{r,i}^2   \\
&+ 2\,C_g \, \dot\varphi_*^{2} + \frac{C_g}8 \, (\dot\varphi_{r,1} - \dot\varphi_{r,2})^2 \biggr). \numberthis
\label{eq:T_two_blocks_final}
\end{align*}

This makes clear that there is a direct capacitive coupling between the two resonator variables as desired, while the unwanted variable $\varphi_*$ decouples. Note that the transformation given in Eq. \ref{eq:eliminate} is nonsingular and that the system variables as defined in Eq. \ref{eq:variables} remain unchanged.
As $\varphi_*$ does not appear at all in the potential energy of the coupled system

\begin{align*}
\mathcal U &= \sum_{i=1}^2  \left(\frac{\Phi_0}{2\pi}\right)^2  \frac{1}{4L_i} (\varphi_{q,i}^2 + \varphi_{r,i}^2) - E_{Jq,i} \cos(\varphi_{q,i}) \\
& - 2E_{J,i} \cos\left(\frac{\varphi_{q,i}}2\right) \sin\left(\frac{\varphi_{r,i}}2\right) \numberthis
\label{eq:pot}
\end{align*}

it can be safely discarded (compare also Appendix \ref{app:cholesky}).\\
Without making any approximations apart from the series expansion in Eq. \ref{eq:series}, we have thus found a system that implements the Hamiltonian proposed by Billangeon (Eq. \ref{eq:long2}).\\
Note that as $\varphi_g$ appears only in the discarded variable $\varphi_*$, a charge offset between ground and the rest of the circuit can never influence our system variables (Eq. \ref{eq:variables}). This means, that even in the whole grid we never need to take charge fluctuations into account, which gives us a lot of freedom in choosing the parameters.

\subsection{Stray capacitances}
\label{sec:stray}

Stray capacitances might appear between a resonator node $\varphi_{c}$ and ground as shown in Fig. \ref{fig:stray}. As we will see, this only leads to a rescaling of the coupling but does not affect its form or its strict locality.

\begin{Figure}
  \begin{center}
    \includegraphics[width=\linewidth]{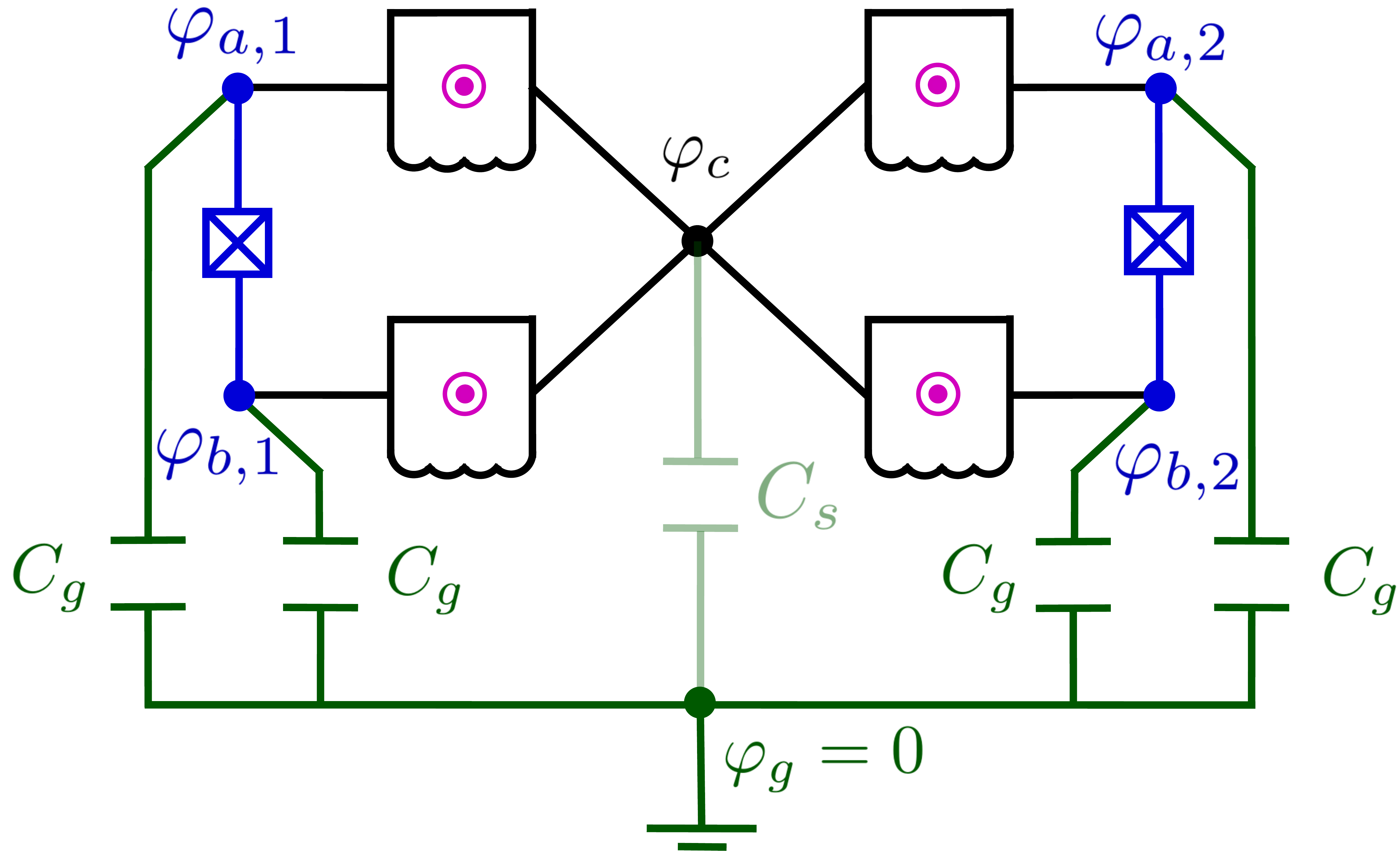}
    \captionof{figure}{A stray capacitance between $\varphi_c$ and ground leads to a rescaling of the coupling.}
    \label{fig:stray}
  \end{center}
\end{Figure}

Such a stray capacitance will add a term $\sim \,C_s (\dot\varphi_c - \dot\varphi_g)^2$ to the kinetic energy given in Eq. \ref{eq:T_two_blocks}. In order to compensate for this extra term, we have to adapt the transformation given in Eq. \ref{eq:eliminate} to

\begin{align}
\varphi_* = \varphi_c - \varphi_g + \frac{C_g}{4 C_g + C_s} (\varphi_{r,1} + \varphi_{r,2})
\label{eq:eliminate_stray}
\end{align}

(this is an application of a Cholesky-decomposition technique; see Appendix \ref{app:cholesky} for the derivation), which finally leads to

\begin{align*}
\mathcal T = \sum_{i=1}^2 & \left(\frac{\Phi_0}{2\pi}\right)^2 \biggl(\frac{2\, C_{q,i} + C_i + C_g}4 \, \dot\varphi_{q,i}^2 + \frac{C_i}4 \, \dot\varphi_{r,i}^2   \\
&+ \frac{4C_g+C_s}2 \, \dot\varphi_*^{2} + \frac{C_g^2}{2(4C_g + C_s)} \, (\dot\varphi_{r,1} - \dot\varphi_{r,2})^2 \biggr). \numberthis
\label{eq:T_two_blocks_stray}
\end{align*}

Note again that Eq. \ref{eq:eliminate_stray} leaves the system variables unchanged. Thus, the only effect of the stray capacitance $C_s$ is the rescaling of the coupling term in Eq. \ref{eq:T_two_blocks_stray}.

\section{Extension to grid}
\label{sec:grid}

\begin{figure*}
 \includegraphics[width=.8\linewidth]{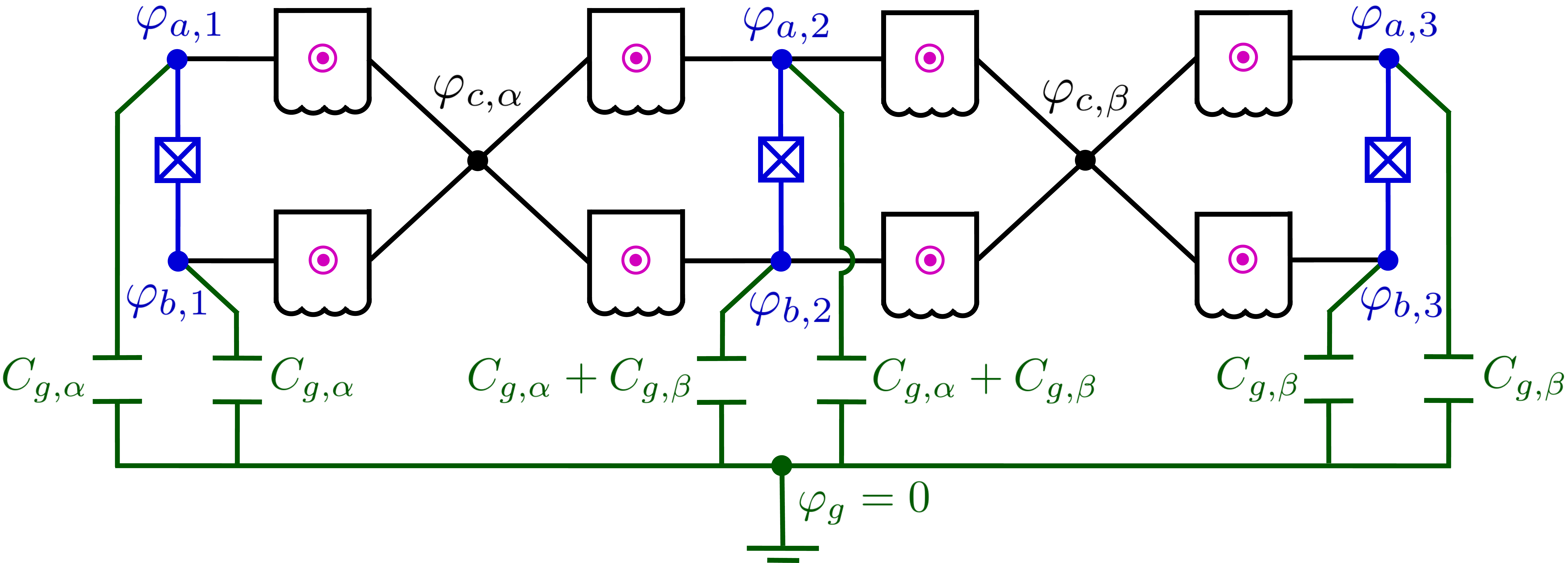}
 \caption{At grid edges, it becomes important that one capacitance to ground is needed for every block a qubit couples to.}
 \label{three_blocks.pdf}
\end{figure*}

It is straightforward to extend this idea to the grid proposed by Billangeon \cite{bill} (see Fig. \ref{grid.pdf}), where every qubit couples longitudinally to its resonators and every resonator couples capacitively to a resonator of the next block. \\
In analogy to the two coupled blocks described in Sec. \ref{sec:two_blocks}, Fig. \ref{ring.pdf} shows a plaquette of four coupled blocks. All qubit nodes (blue colored nodes) are connected to the same ground node $\varphi_g = 0$ (compare Sec. \ref{sec:two_blocks}) via capacitances $C_g$. Note that we need one such capacitance for every connection; that is, a qubit block needs a capacitance to ground for every block it couples to (see Fig. \ref{three_blocks.pdf}). 

\begin{Figure}
  \begin{center}
    \includegraphics[width=.95\linewidth]{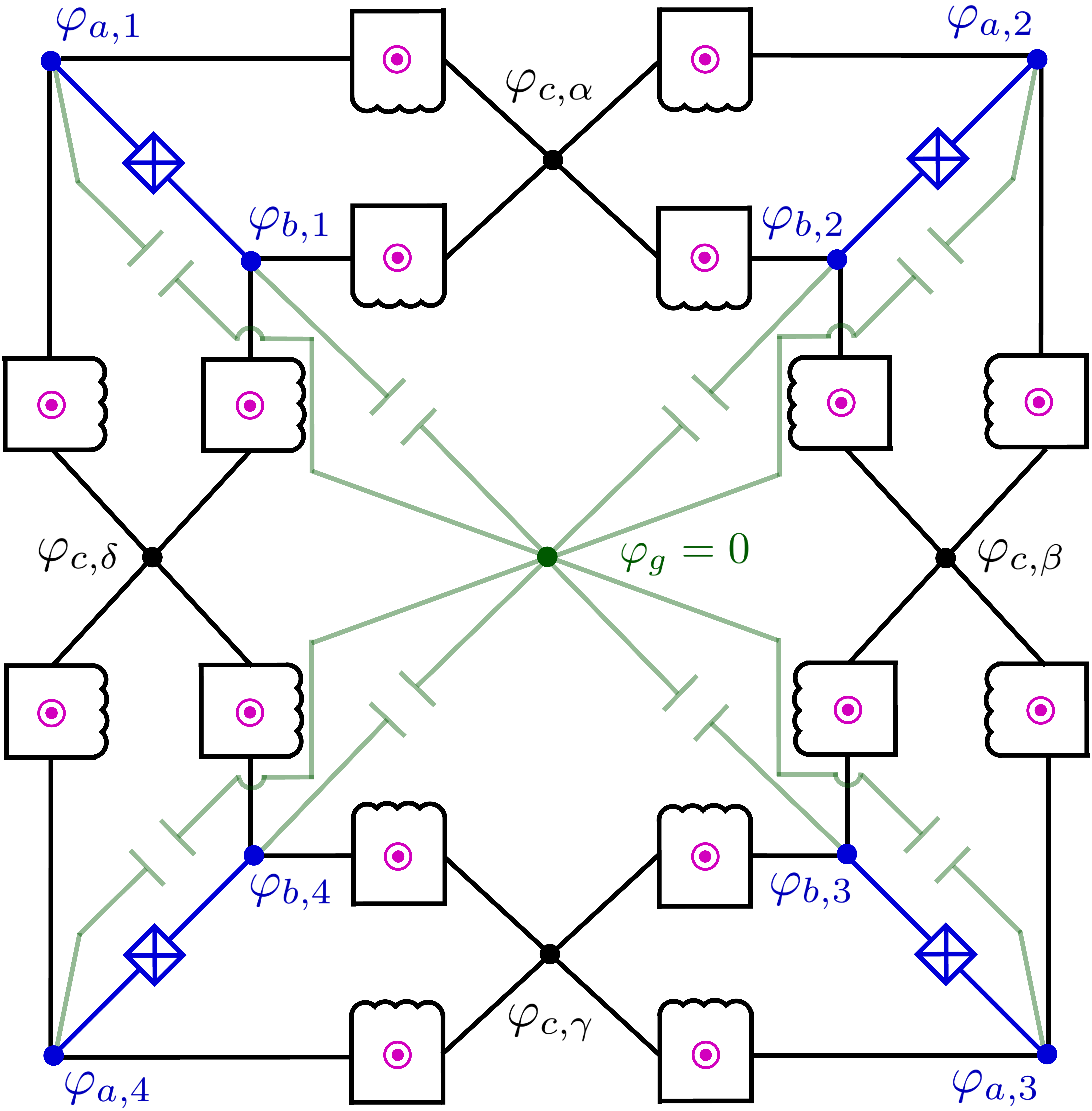}
    \captionof{figure}{This circuits realizes a plaquette of four qubits coupled according to the scheme of \cite{bill}.}
    \label{ring.pdf}
  \end{center}
\end{Figure}

For every unit of two qubits coupled via two resonators, we will find the kinetic energy to be equal to Eq. \ref{eq:T_two_blocks}. In order to emphasize that this coupling is entirely local, we can write the kinetic energy for the ring of four coupled qubits depicted in Fig. \ref{ring.pdf}. It is

\begin{align}
\mathcal T \hspace{-0.05cm}=\hspace{-0.05cm} \sum_{i,j} \left(\frac{\Phi_0}{2\pi}\right)^2 \hspace{-0.1cm} \biggl(\frac{2 C_{q,i} + C_{i,j} + C_{g,j}}2 \, \dot\varphi_{q,i}^2 +\frac12 \, \dot{\boldsymbol\varphi}_j^T \mathbf{C}_j \, \dot{\boldsymbol\varphi}_j \hspace{-0.05cm}\biggr),
\end{align}

with $\boldsymbol{\varphi}_j^T = (\varphi_{c,j}, \varphi_{r,1,j}, \varphi_{r,2,j})$ as defined in Eq. \ref{eq:variables}, where $\varphi_{r,1,j}$ and $\varphi_{r,2,j}$ are the two resonators connected at node $\varphi_{c,j}$, and

\begin{align}
\mathbf{C}_j =\begin{pmatrix}
4\, C_{g,j} & C_{g,j} & C_{g,j} \\
C_{g,j} & \frac{C_{1,j} + C_{g,j}}2 & 0 \\
C_{g,j} & 0 & \frac{C_{2,j} + C_{g,j}}2 
\end{pmatrix},
\end{align}

where $i = 1,2,...$ numbers the qubits and $j = \alpha, \beta,... $ numbers the connections between two coupled blocks as depicted in Fig. \ref{ring.pdf}.
It becomes clear that this is a completely local problem, as there is no coupling at all between $\mathbf{C}_\alpha$ and $\mathbf{C}_\beta$. Introducing 

\begin{align}
\varphi_{*,j} = \varphi_{c,j} + \frac{\varphi_{r,1,j} + \varphi_{r,2,j}}{4}
\end{align}

(compare Eq. \ref{eq:eliminate}) for each of these connections, we can independently decouple the superfluous variables $\varphi_{*,j}$ and discard them (see also Appendix \ref{app:cholesky}).
We find that for every connection $j$ between two coupled blocks, we can always write the kinetic energy as

\begin{align*}
\mathcal T_j &= \sum_{i=1}^2  \left(\frac{\Phi_0}{2\pi}\right)^2 \biggl(\frac{2\, C_{q,i} + C_{i,j} + C_{g,j}}4 \, \dot\varphi_{q,i}^2    \\
&+ \frac{C_{i,j}}4 \, \dot\varphi_{r,i,j}^2+ 2\,C_{g,j} \, \dot{\varphi}_{*,j}^2 + \frac{C_{g,j}}8 \, (\dot\varphi_{r,1,j} - \dot\varphi_{r,2,j})^2 \biggr) \numberthis
\end{align*}

(which is equal to Eq. \ref{eq:T_two_blocks_final}). \\
In total, we can say that while the potential energy contains the longitudinal coupling between a qubit and its resonators with no connection between two blocks 1 and 2, the kinetic energy contains the transverse coupling between two resonators, with no connection between two such units $\alpha$ and $\beta$.

\section{Driving the qubit for single and two-qubit gates}

\begin{wrapfigure}{r}{0.7\linewidth}
\vspace{-20pt}
  \begin{center}
    \includegraphics[width=.85\linewidth]{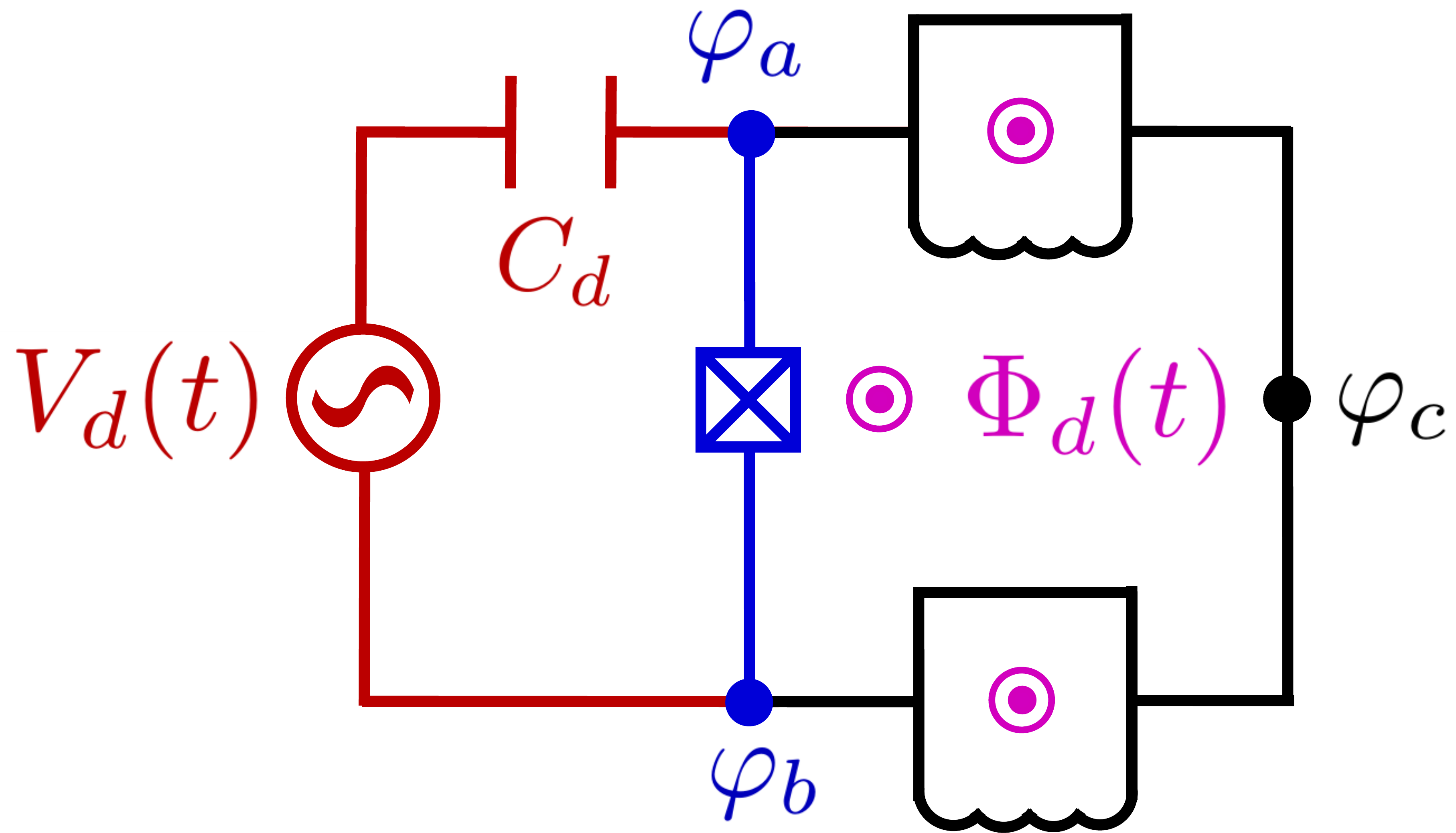}
  \end{center}
  \vspace{-10pt}
   \caption{The qubit can be driven by either connecting a voltage source to the qubit nodes $\varphi_a$ and $\varphi_b$ or by applying a flux through the main loop.}
    \label{driving.pdf}
\end{wrapfigure}

Read-out and gates can be performed by applying a transverse microwave drive on the qubit. Connecting an AC voltage source to the qubit nodes $\varphi_a$ and $\varphi_b$ as shown in Fig. \ref{driving.pdf} yields, with $\dot\varphi_d = \frac{2\pi}{\Phi_0} V_d (t)$,

\begin{align*}
\mathcal L_d &= \left(\frac{\Phi_0}{2\pi}\right)^2 \frac{C_d}{2} (\dot\varphi_d - (\dot\varphi_a - \dot\varphi_b))^2 \\
&=  \frac{C_d}{2} V_d(t)^2 - \frac{\Phi_0}{2\pi} C_d V_d(t) \dot \varphi_q + \left(\frac{\Phi_0}{2\pi}\right)^2 \frac{C_d}{2} \dot\varphi_q^2. \numberthis
\end{align*}

While the first term is a scalar offset that will be ignored, the second term corresponds to a transverse driving term $\sim A \cos(\omega t)\, \sigma_x$ as it is an odd function in $\varphi_q$ (compare Sec. \ref{sec:long}). The third term adds to the kinetic energy of the qubit.  \\
Alternatively, a small flux $\Phi_d (t) = \Phi_0/(2\pi) \, A \cos(\omega t)$ through the main loop would lead to the same driving term, as

\begin{align}
\cos\left(\varphi_q - A \cos(\omega t) \right) \approx \cos(\varphi_q) +  A \cos(\omega t) \sin(\varphi_q)
\end{align}

for $A \ll \varphi_q$, which corresponds again to a $\sigma_x$-type transverse drive, as $\sin(\varphi_q)$ is an odd function in $\varphi_q$ (compare Sec. \ref{sec:long}). The first term adds to the potential energy of the qubit. 

Transforming this drive to the frame where $\mathcal H$ is diagonal (compare Eq. \ref{eq:lang-firsov}), it can be used to drive both single-qubit and qubit-resonator sideband transitions.
As shown explicitly in \cite{bill}, single-qubit rotations can be done by driving the qubit at its transition frequency $\omega = \Delta$. 
Similarly, sideband transitions between the qubit and both the nearest and next-nearest 
resonator can be driven by choosing the drive frequency to be $\omega = |\Delta \pm \Omega_{r,1} \pm \Omega_{r,2}|$ (compare Appendix \ref{app:diagonal}). As shown in \cite{bill}, a series of three of these sideband transitions leads to a 
two-qubit phase gate.

%Similarly, sideband transitions between the qubit and both the nearest and next-nearest resonator can be driven by choosing the drive frequency to be $\omega = |\Delta \pm \omega_{r,1} \pm \omega_{r,2}|$. Consulting Fig. \ref{grid.pdf}, we see that it suffices to make all of the quantities $\pm \omega_1 \pm \omega_2$, $\pm \omega_1 \pm \omega_4$, $\pm \omega_3 \pm \omega_2$, and $\pm \omega_3 \pm \omega_4$ distinct and nonzero. 
%As shown in \cite{bill}, a series of four of these sideband transitions leads to a two-qubit phase gate.

\section{Summary}

We presented a proposal for a circuit QED system where a qubit couples to resonators via its longitudinal degree of freedom and every resonator is capacitively coupled to a resonator from the next unit cell. This proposal is easily scalable to any number of resonators per qubit and any number of unit cells as shown in Secs. \ref{sec:n_resonators}-\ref{sec:grid}. \\
This constitutes an implementation of an idea presented by Billangeon et al. in \cite{bill}. As they show explicitly, such a system is exactly diagonalizable using a series of unitary transformations. In this diagonal frame, there are no dispersive shifts or residual couplings between any qubits or resonators.
Single-qubit operations and sideband transitions between a qubit and any of its resonators can be done by driving the qubit at a certain frequency. The coupling is strictly confined to nearest and next-nearest neighbor resonators of each qubit; there is never any direct qubit-qubit coupling.\\
Our proposal for the qubit-resonator system should be inherently unaffected
by charge fluctuations, which leaves us a lot of freedom to choose our system parameters. In particular, the longitudinal coupling between the qubit and a resonator can be tuned independently of the other parameters (see Appendix \ref{app:hamiltonian}).

\section*{Acknowledgments}
We acknowledge support from the Alexander von Humboldt foundation and from ScaleQIT.

\end{multicols}

\vspace{.75cm}

%\hrulefill

\vspace{.75cm}

\begin{multicols}{2}
\appendix

\section{Diagonalization}
\label{app:diagonal}

As shown explicitly in \cite{bill}, the Hamiltonian

\begin{align*}
\mathcal{H} = \sum_{i=1}^2 & \omega_i\, a_i^\dagger a_i +  \frac{\Delta_i}{2} \, \sigma_{i}^z + g_i \, \sigma_i^z \, (a_i^\dagger + a_i)  \\
&- g_c \,(a_1^\dagger - a_1)(a_2^\dagger - a_2), \numberthis
%\label{eq:long2}
\end{align*}

can be exactly diagonalized by a series of unitary transformations leading to

\begin{align}
\mathcal{H}' = \Omega_+ a_1^\dagger a_1 + \Omega_- a_2^\dagger a_2 + \sum_{i=1}^2 \frac{\Delta_i}{2} \sigma_i^z - \left(\frac{g_1^2}{\omega_1} + \frac{g_2^2}{\omega_2}\right) \mathbf{1}.
\end{align}

We can see that while the qubit frequencies are unaffected by the transformations, the resonator frequencies get shifted to

\begin{align}
\Omega_\pm² = \frac{\omega_1²+\omega_2²}{2} \pm \frac{1}{2} \sqrt{(\omega_1²-\omega_2²)²+16 \,g_c² \omega_1 \,\omega_2}.
\end{align}

A transverse drive on one of the qubits

\begin{align}
\mathcal{H}_d(t) = \Omega \cos(\omega t + \phi) \sigma_1^x
\end{align}

has to be transformed to the same frame. In a rotating frame, where the qubits and the resonators stand still, we can then apply the rotating wave approximation and frequency-select the gate we want to drive. While single-qubit operations are implemented at $\omega = \Delta_1$, sideband transitions between the qubit and either one or both neighboring resonators can be driven using $\omega = |\Delta_1 \pm \Omega_\pm|$ or $\omega = |\Delta_1 \pm \Omega_+ \pm \Omega_-|$, respectively. On a grid, where every qubit has four nearest neighbor resonators and four next-nearest neighbor resonators (see Fig. \ref{grid.pdf}), we need eight different frequencies $\Omega_\pm^i$, $i=1,...,4$, for unequivocal frequency selection. Remarkably, as $\Omega_+ \neq \Omega_-$ for $g_c \neq 0$, this can be realized even if all eight original frequencies were equal, just by having four different values of $g_c$.

\section{Unequal junctions}
\label{app:unequal}

The characteristics of the qubit proposed in Sec. \ref{sec:qubit} highly depend on the fact that the upper and lower resonator branch (see Fig. \ref{qubit_resonator.pdf}) are exactly identical. While it is relatively easy to build two equal capacitances or inductances, this is quite difficult for Josephson junctions. Suppose the two junctions have slightly different Josephson energies $E_{J_1}$ and $E_{J_2}$, then the coupling term is

\begin{align*}
&E_{J,1} \sin\left(\frac{\varphi_r + \varphi_q}2\right) + E_{J,2} \sin\left(\frac{\varphi_r - \varphi_q}2\right)\\
&= E_{J\Sigma} \biggl(\sin\left(\frac{\varphi_r}2\right) \cos\left(\frac{\varphi_q}2\right) + d \cos\left(\frac{\varphi_r}2\right) \sin\left(\frac{\varphi_q}2\right) \biggr), \numberthis
\label{eq:unequal}
\end{align*}

where $d= (E_{J,2} - E_{J,1})/E_{J\Sigma}$ is the junction asymmetry and $E_{J\Sigma} = E_{J,1} + E_{J,2}$ (compare \cite{didier}). The second term in Eq. \ref{eq:unequal} is an unwanted transverse coupling term $\sim \sigma_x \, a^\dagger a$, that cancels out if $E_{J_1} = E_{J_2}$. As a remedy to this problem, one could consider making one or both of the junctions tunable by using a  SQUID (superconducting quantum interference device). Note that the qubit and the resonator are completely uncoupled for $E_{J,1} = E_{J,2} = 0$. Two SQUIDs instead of the two junctions would therefore lead to tunable coupling. \\
Equivalently, transverse coupling terms arise if the inductances or capacitances in the upper and lower resonator branch are not identical. However, they will always have the form $\sigma_x (a^\dagger \pm a)$ and can therefore not compensate the term mentioned in Eq. \ref{eq:unequal}.

\section{Hamiltonian formulation}
\label{app:hamiltonian}

We can rewrite the qubit Lagrangian (Eq. \ref{eq:lagr_final}) in the Hamiltonian formulation. For $E_J = 0$, the coupling to the resonator is zero and the Lagrangian for the qubit only is

\begin{align}
\mathcal L_q &= \left(\frac{\Phi_0}{2\pi}\right)^2 \left(\frac{C_{\text{tot}}}4 \, \dot\varphi_q^2 - \frac{1}{4L_{\text{tot}}} \, \varphi_q^2 \right) + E_{Jq} \cos(\varphi_q)
\end{align}

with $C_{\text{tot}} = 2\,C_q + \sum_{j} (C_{j} + C_{g,j})$ and $1/L_{\text{tot}} = \sum_{j} 1/L_j$, where $C_j$ and $L_j$ are the capacitances and inductances of the resonator arms connected to the qubit and $C_{g,j}$ are the capacitances to ground (compare Sec. \ref{sec:grid}).\\ 
The conjugate variable to the phase $\varphi_q$ is the rescaled charge $n = Q/(2e)$ (number of Cooper pairs) with

\begin{align}
n = \frac{1}{\hbar}\frac{\partial \mathcal L}{\partial \dot\varphi_q}.
\end{align}

We thus find $ \dot\varphi_q = 2\,n \, \hbar /C_{\text{tot}} (2\pi/\Phi_0)^2$, and with $E_{C} =e^2/(2C_{\text{tot}})$ and $\Phi_0 = \pi \hbar/e$ the Hamiltonian yields

\begin{align*}
\mathcal H_q = 8 E_{C}\,  n^2 + \frac 1{4 L_{\text{tot}}} \left(\frac{\Phi_0}{2\pi}\right)^2  \varphi_q^2 -E_{Jq} \cos(\varphi_q). \numberthis
\end{align*}

Using the rescaled Josephson energy

\begin{align}
E_{Jq}^* = E_{Jq} + \frac1{2L_{\text{tot}}} \left(\frac{\Phi_0}{2\pi}\right)^2,
\end{align}

the Hamiltonian can be written as

\begin{align*}
\mathcal H_q = 8 E_{C}\, &n^2 -E_{Jq} + \frac{1}{2} E_{Jq}^* \varphi_q^2 - \frac1{24} E_{Jq} \varphi_q^4  \numberthis
\end{align*}

up to fourth order in $\varphi_q $. Now we quantize, treating $\hat n \sim i(c^\dagger - c)$ and $\hat\varphi_q \sim c^\dagger + c$ as operators with $c = \sum_m \sqrt{m+1} |m\rangle\langle m+1|$ and $[\varphi_q,n] = i$ and write

\begin{align}
\mathcal H_q = \Delta \, c^\dagger c + \frac{\delta}{12} (c^\dagger + c)^4,
\label{eq:ham}
\end{align}

where

\begin{align}
\Delta = 4 \sqrt{E_{Jq}^* E_{C}}
\end{align}

is the gap of the qubit and 

\begin{align}
\delta = - \frac{2 E_{C} E_{Jq}}{E_{Jq}^*}
\end{align}

is its anharmonicity (compare \cite{koch}). If the anharmonicity is large enough, we can treat the qubit as a two-level system and write

\begin{align}
\mathcal{H}_q = \frac{\Delta}{2} \sigma_z
\end{align}

instead of Eq. \ref{eq:ham}.
As charge fluctuations do not matter here, we have a lot of freedom in choosing these parameters. Notably, there is no reason for the anharmonicity to be small.\\
In the same way, we can find expressions for the resonator frequencies and couplings in a coupled system of two qubits and two resonators (as described in Sec. \ref{sec:two_blocks}). Using $\hbar = 1$, the frequency of the first resonator yields

\begin{align}
\omega_{r,1} = \sqrt{\frac{2C_2 + C_g}{(2C_1 C_2 + C_g (C_1 + C_2))L_1}}
\end{align}

and similarly for $\omega_{r,2}$, while we find 

\begin{align}
g_1 = - \frac{E_{J,1}}2 \sqrt{\frac{\pi^3}{e \, \Phi_0^3}} \left(\frac{(2C_2 + C_g)L_1}{2C_1 C_2 + C_g (C_1 + C_2)}\right)^{\frac14}\sqrt{\frac{E_{C,1}}{E_{Jq,1}^*}}
\end{align}

for the longitudinal coupling between qubit 1 and its resonator and 
\begin{small}
\begin{align}
g_c = \frac{\hbar \, C_g}{2 \sqrt{2C_1 C_2 + C_g (C_1 + C_2)}((2C_1+C_g)(2C_2+C_g)L_1 L_2)^{\frac14}}
\end{align}
\end{small}

for the capacitive coupling between the two resonators. Note that $E_{J_i}$ only appears in the term for the longitudinal coupling. We are thus able to tune this coupling independently of the other variables. Likewise, the Josephson energy of the qubit junction $E_{Jq}$ appears only in the expression for the qubit gap $\Delta$ and its anharmonicity $\delta$. If we choose $C_g$ to be large to achieve strong coupling between the resonators, we can still have a high anharmonicity as long as $E_{Jq}$ is large enough.

\section{Variable elimination using the Cholesky transformation}
\label{app:cholesky}

The so-called \textit{Cholesky} decomposition of a Hermitian positive-definite matrix $\mathbf A$ consists of an upper triangular matrix $\mathbf B$ with real and positive diagonal entries and its conjugate transpose $\mathbf B^\dagger$, that is,

\begin{align}
\mathbf A = \mathbf{B}^\dagger \mathbf{B}.
\end{align}

This decomposition has oftentimes a much simpler form than a square root decomposition, making analytic calculations a lot easier. 
As we will see, the Cholesky transformation can be used for the elimination of unwanted variables (see Secs. \ref{sec:two_blocks} and \ref{sec:stray}).\\
Let us consider the kinetic energy for two coupled blocks given in Sec. \ref{sec:two_blocks} (Eq. \ref{eq:T_two_blocks}) including the stray capacitance term introduced in Sec. \ref{sec:stray}, that is,

\begin{align}
\mathcal T = \sum_{i} \left(\frac{\Phi_0}{2\pi}\right)^2 \biggl(\frac{2\,C_{q,i} + C_{i} + C_{g}}2 \, \dot\varphi_{q,i}^2 +\frac12 \, \dot{\boldsymbol\varphi}^T \mathbf{C} \, \dot{\boldsymbol\varphi} \biggr),
\end{align}

with $\boldsymbol{\varphi}^T = (\bar\varphi = \varphi_c - \varphi_g, \varphi_{r,1}, \varphi_{r,2})$ as defined in Eq. \ref{eq:variables}, and

\begin{align}
\mathbf{C} =\begin{pmatrix}
4 C_{g} + C_s & C_{g} & C_{g} \\
C_{g} & \frac{C_{1} + C_{g}}2 & 0 \\
C_{g} & 0 & \frac{C_{2} + C_{g}}2 
\end{pmatrix}.
\end{align}

While it is not trivial to find the eigenvalues and eigenvectors of $\mathbf{C}$, its Cholesky decomposition (the first row in particular) has a quite simple form:

\begin{small}
\begin{align}
\mathbf{B} = \begin{pmatrix}
\sqrt{4 C_g + C_s} & \frac{C_g}{\sqrt{4 C_g + C_s}} & \frac{C_g}{\sqrt{4 C_g + C_s}} \\
0 & \sqrt{\frac{C_g (2C_g + C_s)}{2 (4 C_g+C_s)} + \frac{C_1}{2}} & ... \\
0 & 0 & ...
\end{pmatrix}
\end{align}
\end{small}

(we only need the first row here). 
We would like to decouple $\bar\varphi = \varphi_c - \varphi_g$ from the resonator variables $\varphi_{r,1}$ and $\varphi_{r,2}$. As $\bar\varphi$ does not appear in the potential energy (Eq. \ref{eq:pot}), we can basically transform it to any combination of the three variables without changing $\mathcal U$. Consider therefore the transformation

\begin{align}
\mathbf{R} =\begin{pmatrix}
1 & \frac{C_g}{4 C_g + C_s} & \frac{C_g}{4 C_g + C_s} \\
0 & 1 & 0 \\
0 & 0 & 1 
\end{pmatrix},
\end{align}

which consists of the first row of the Cholesky decomposition of $\mathbf{C}$ (rescaled to be unitless) and an identity matrix for the other two rows. It is clear that the transformation $\mathbf{R} \, \boldsymbol{\varphi}$ leaves the resonator variables $\varphi_{r,1}$ and $\varphi_{r,2}$ untouched and that therefore the potential energy is unchanged.
The kinetic energy, however, transforms to

\begin{align}
\mathcal T = \sum_{i} \left(\frac{\Phi_0}{2\pi}\right)^2 \biggl(\frac{2\,C_{q,i} + C_{i} + C_{g}}2 \, \dot\varphi_{q,i}^2 +\frac12 \, \boldsymbol{\dot{\tilde\varphi}}^T \mathbf{\tilde{C}} \, \boldsymbol{\dot{\tilde\varphi}} \biggr)
\end{align}

with

\begin{small}
\begin{align}
\mathbf{\tilde{C}} = \begin{pmatrix}
4 C_{g} + C_s & 0 & 0 \\
0 & \frac{C_1}{2} + \frac{C_g (2C_g + C_s)}{2 (4C_g + C_s)} & - \frac{C_{g}^2}{4 C_g + C_s} \\
0 & - \frac{C_{g}^2}{4 C_g + C_s} & \frac{C_1}{2} + \frac{C_g (2C_g + C_s)}{2 (4C_g + C_s)} 
\end{pmatrix}
\end{align}
\end{small}

for $\boldsymbol{\tilde\varphi}^T = (\varphi_*, \varphi_{r,1}, \varphi_{r,2})$, which is equal to the results given in Eqs. \ref{eq:T_two_blocks_final} (for $C_s = 0$) and \ref{eq:T_two_blocks_stray}. Without changing the resonator variables $\varphi_{r,1}$ and $\varphi_{r,2}$, we have thus gone to a frame where the unwanted variable 

\begin{align}
\varphi_* = \varphi_c - \varphi_g + \frac{C_g}{4 C_g + C_s} (\varphi_{r,1} + \varphi_{r,2})
\end{align}

is uncoupled and can therefore be discarded. \\
As the first row of a Cholesky decomposition is usually very simple, this method can be easily used in such cases, where the kinetic energy includes more variables than the potential energy (or vice versa).

\end{multicols}

\vspace{.75cm}

\hrulefill

\vspace{.75cm}

\begin{multicols}{2}

\nocite{*} % Insert publications even if they are not cited in the poster
\small{\bibliographystyle{unsrt}
\bibliography{Bibliography}\vspace{0.75in}}

\end{multicols}
\end{document}